\documentclass{epl}
\title{Quantum and classical echoes in scattering systems described by 
simple Smale horseshoes}
\shorttitle{Quantum and classical echoes in scattering systems}
\author{C. Jung\inst{1} \and C. Mejia-Monasterio\inst{1} \and T.H. 
Seligman\inst{1,2}}
\shortauthor{C. Jung \etal}
\institute{
  \inst{1}Centro de Ciencias F\'\i sicas, University of Mexico (UNAM), 
  Cuernavaca, Mexico\\
  \inst{2} Centro internacional de Ciencias, Cuernavaca, Mexico
}
\pacs{05.45.Mt}{Semiclassical chaos (``quantum chaos'')}
\pacs{03.65.Nk}{Scattering theory}

\begin{document}

\maketitle

\begin{abstract}
We explore the quantum scattering of systems classically described by
binary and other low order Smale horseshoes, in a stage of development
where the stable island associated with the inner periodic orbit is
large, but chaos around this island is well developed.  For short
incoming pulses we find periodic echoes modulating an exponential
decay over many periods. The period is directly related to the
development stage of the horseshoe. We exemplify our studies with a
one-dimensional system periodically kicked in time and we mention
possible experiments.
\end{abstract}

In classical mechanics the Smale horseshoe \cite{smale} construction
has proven to be the key point to the understanding of chaotic
scattering in time independent systems with two degrees of freedom and
time dependent ones with one degree of freedom \cite{jung-1,rueckerl}.
Though the importance of this construction in the quantum analogue of
such systems has been noticed occasionally \cite{borondo}, a study of
the implications in quantum systems has not yet been undertaken.  We
are interested in low-order (such as binary or ternary) horseshoes,
where the features encountered are comparatively simple. In the
present paper we shall concentrate on situations in which the stage of
development of the horseshoe is fairly low. We will discuss a binary
horseshoe for which one of the fundamental periodic orbits is
hyperbolic and shows homoclinic connections, while the other one is
still elliptic and confined inside a large stable island. In such a
situation tunneling into the island will be the most notable quantum
effect. We shall show that a short pulse as incoming wave leads to
periodic pulses in the outgoing wave. They survive many periods and we
shall call them echoes. If we use good energy resolution instead we
find narrow resonances.

We will focus our discussion on a one-dimensional kicked scattering
model mainly because of the ease of calculation, but experiments with
similar periodically driven models may be of interest
\cite{raizen}. Yet the effect is not confined to such
models. Two-dimensional time-independent billiards with two openings
or leads can produce ternary horseshoes with a large central island,
whose echoes may be seen in microwave experiments \cite{richter} or
mesoscopic systems.

The model we use is given in terms of the Hamiltonian
\begin{equation} \label{hamiltonian}
H(q,p,t) = \frac{p^2}{2} + A \ V(q) \sum_{n=-\infty}^{\infty} \delta (t - 
n) \ .
\end{equation}
The time dependence is an infinite periodic train of delta pulses
kicking the potential with period $1$.  The parameter $A$ determines
the strength of the potential, which is given by
\begin{equation} \label{potential}
V(q) = \left\{
\begin{array}{ccl}
\frac{\textstyle q^2}{\textstyle 2} + 1 &,& q < 0 \\
\\
e^{-q}(q^2 + q + 1) &,& q \geq 0 \\
\end{array}
\right..
\end{equation}
Note that in the literature \cite{rueckerl,gaspard} a similar but
simpler potential is used, where the exponential form extends to
negative values of $q$. We have changed the form of the potential to
obtain better convergence of the quantum calculation. Yet the
classical results for the two potentials are quite similar. We obtain
a smooth development of the binary horseshoe as a function of the
parameter $A$. The only difference is that for the present potential
the development is not entirely monotonic \cite{thesis}, but this does
not affect our considerations.

We represent the classical dynamics by a Poincar\'e map which we
choose as the stroboscopic map taken at times $t = n + 1/2$

\begin{equation} \label {stroboscopic-map}
\begin{array}{lcl}
p_{n+1} & = & p_n - AV'(q_n+p_n/2) \\
q_{n+1} & = & q_n + p_n - A V'(q_n+p_n/2)/2 \ .
\end{array}
\end{equation}
It gives us the evolution of a classical trajectory from time $n+1/2$
to time $n+3/2$.

The kicked potential has a maximum at $q=1$ and a minimum at $
q=0$. It is quite obvious that the points $p=0,\, q=0$ and $q=1,\,
p=0$ are fixed points of this map. Indeed they represent the
fundamental periodic orbits, that determine the construction of the
binary horseshoe, which describes the topology of the Hamiltonian flow
of our system.  The fixed point at $q=1$ is obviously hyperbolic,
while the other one can vary according to the strength parameter
$A$. For $A<4$ this point is elliptic. At this value it turns inverse
hyperbolic, but many secondary islands of stability survive. The phase
portrait is hyperbolic when the horseshoe becomes complete near
$A=6.25$. Hyperbolic stages are also possible for smaller values
of $A$ for which the horseshoe is incomplete \cite{rueckerl,davis,troll}.

\begin{figure}
\onefigure[scale=0.4]{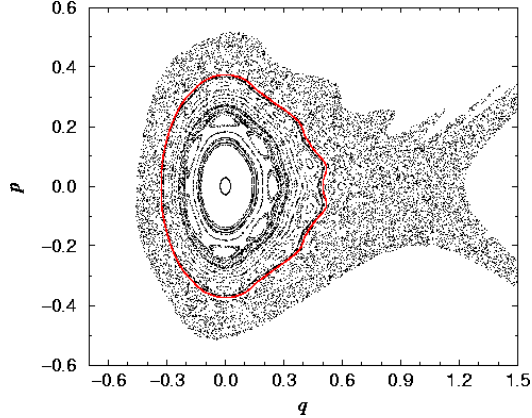}
\caption{Phase portrait of the stroboscopic map
eq.~(\ref{stroboscopic-map}), for $A=0.967$. Together with the bounded
orbits, trajectories corresponding to asymptotic incoming initial
conditions are plotted as well. The thick red line corresponds to the
outermost KAM surface.}
\label{fig-1}
\end{figure}

We shall focus our attention on the low development stages long before
the original elliptic orbit bifurcates.  On the other hand we wish to
see a well developed chaotic region.  These conditions are met for
values of $A$ between 0.5 and 3.  We shall use the parameter
$A=0.967$. A phase portrait of the stroboscopic map for this value is
shown in Fig.~\ref{fig-1}. This value of $A$ was chosen because
somewhere between this value and $A=1$, the outermost KAM surface
shown in Fig.~\ref{fig-1} disintegrates.

We are interested in learning about the quantum properties of such a
low order horseshoe with a large stable island.  For this purpose we
use the unitary time evolution operator, which is rather easily
obtained for kicked systems.  This is the case because the
stroboscopic map eq.~(\ref{stroboscopic-map}) can be decomposed into
three transformations as follows:

\begin{eqnarray}
\label{step1}
p_{n'} & = & p_n \nonumber\\
q_{n'} & = & q_n + p_n/2 \\
\label {step2}
p_{n''} & = & p_{n'} - AV'(q_{n'}) \nonumber\\
q_{n''} & = & q_{n'} \\
\label {step3}
p_{n+1} & = & p_{n''} \nonumber\\
q_{n+1} & = & q_{n''} + p_{n''}/2 \ .
\end{eqnarray}

The second step (\ref{step2}), can be interpreted as a gauge
transformation in coordinate space and the other two (\ref{step1},
\ref{step3}), as a gauge transformations in momentum space.  This
implies, that the unitary operator for one time step can be written as
three phases intertwined by Fourier transforms $\mathcal{F}$, which
take us from coordinate to momentum space and back.  Thus we obtain
for the kernel of this operator in momentum-space
\begin{equation}
\label {u-time-evolution} U (p',p) = \exp\Big[{-\frac{i}{4\hbar} \,
{p^2}}\Big] \ \mathcal{F} \ \exp \Big[\frac{i}{\hbar} \, AV(q)\Big] \
\mathcal{F}^{-1} \, \exp\Big[{-\frac{i}{4\hbar} \, {p^2}}\Big] \ .
\end{equation}
This expression is simple as it involves only Fourier transforms and
multiplication with phases. It is also very efficient if good fast
Fourier transform (FFT) codes are used.

We shall analyze our scattering system in terms of wave packet
dynamics. In all our simulations we use minimum uncertainty Gaussian
wave packets given by

\begin{equation} \label {wave-packet}
\Psi(q,0) = \frac{1}{\pi^{1/4}\sigma^{1/2}} 
\exp \Big[{-\frac{(q-q_{\rm in})^2}{2\sigma^2} + \frac{i}{\hbar}p_{\rm in} 
q}\Big] \ .
\end{equation}

For a given value of the initial momentum $p_{\rm in}$, $\sigma$ will
determine the duration of the pulse and the value of $\hbar$ will
determine how near to the classical limit we operate. Recall that
short pulses imply a poor energy resolution, while long ones can have
very well defined energies.  We can therefore use short pulses and
consider the time evolution in configuration space or in phase space,
or we can use long pulses and look at the energy dependence of some
outgoing quantity.  We shall start with the former and then consider
the latter.

\begin{figure}
\twoimages[scale=0.35]{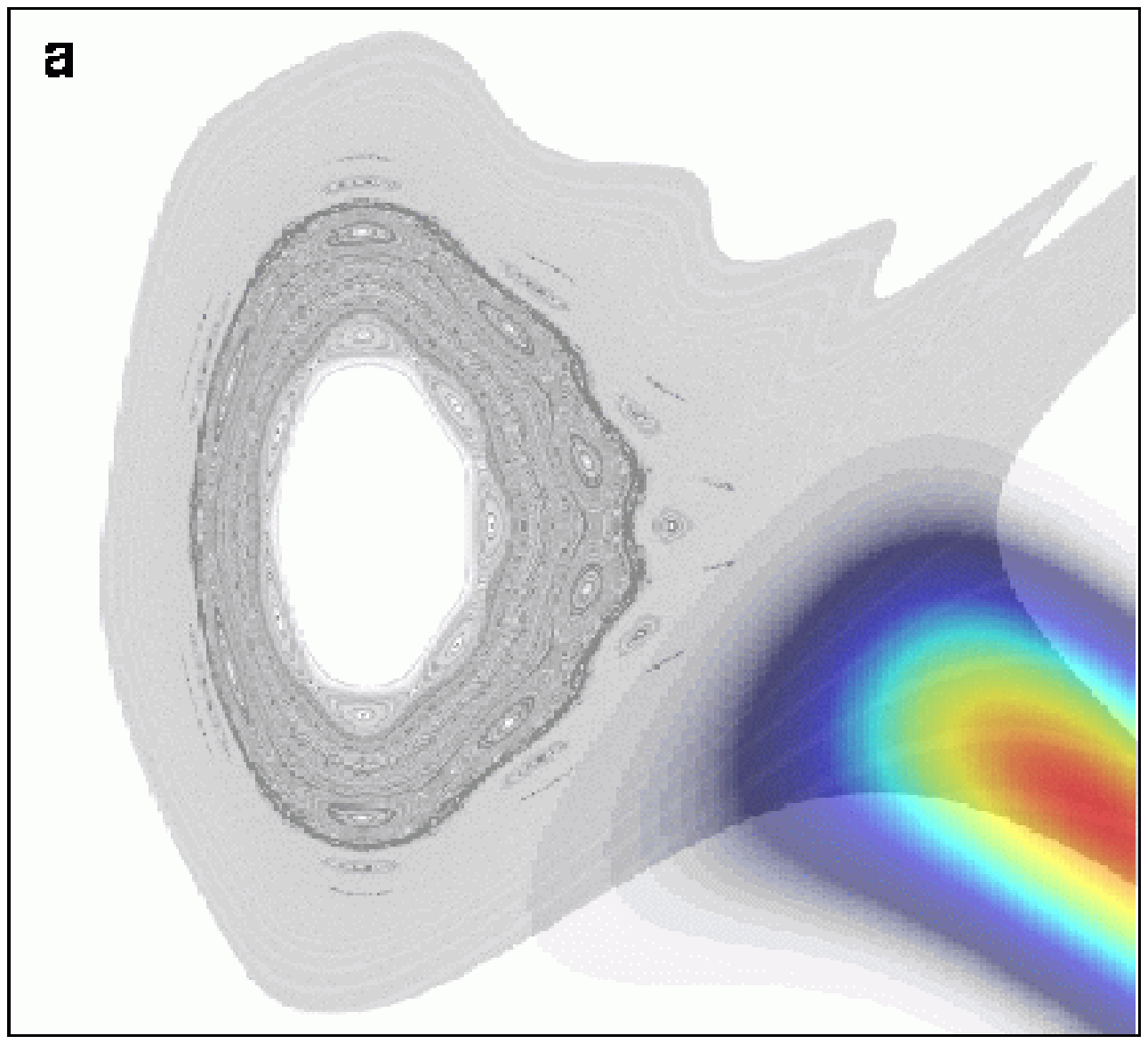}{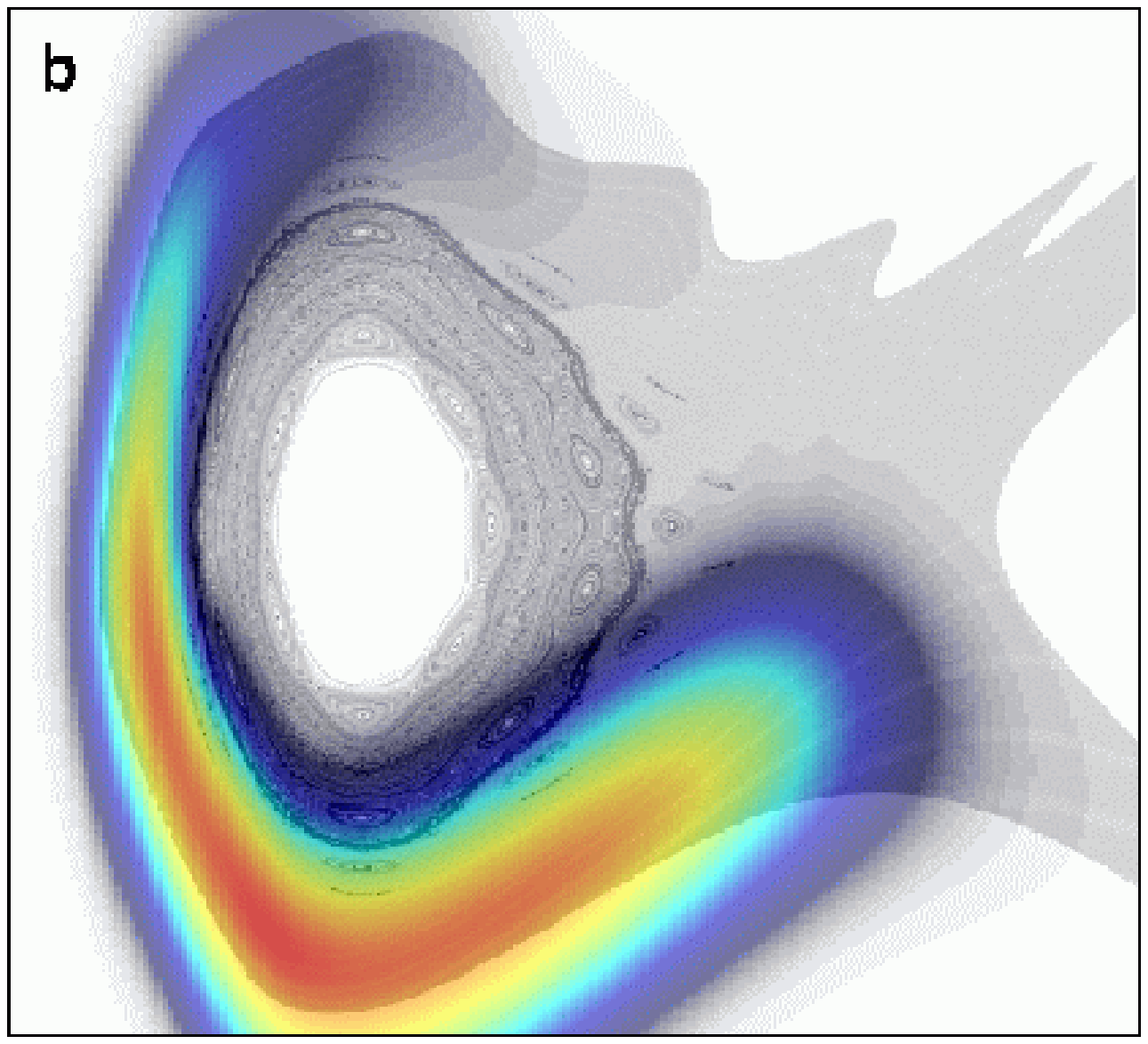}
\twoimages[scale=0.35]{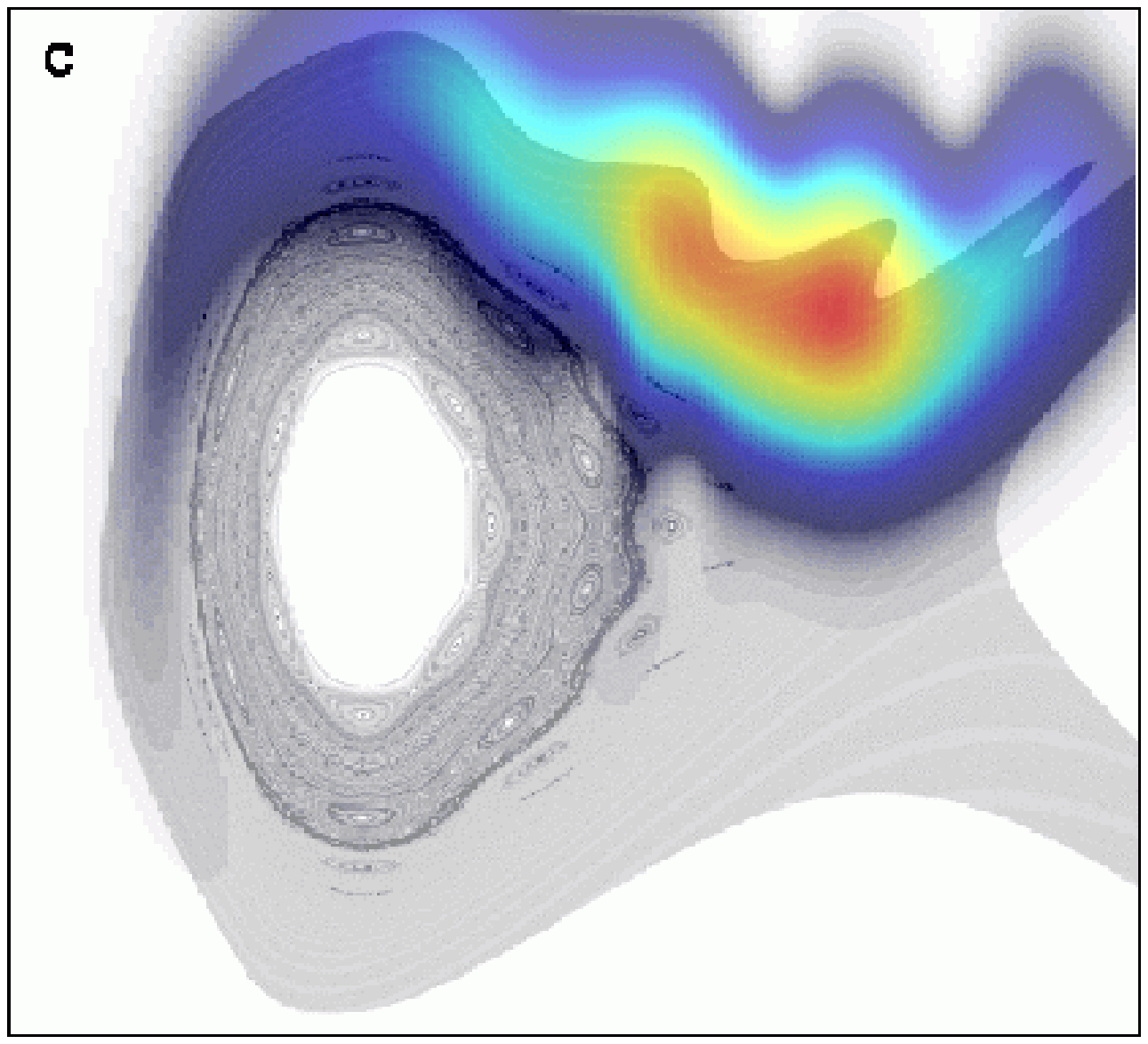}{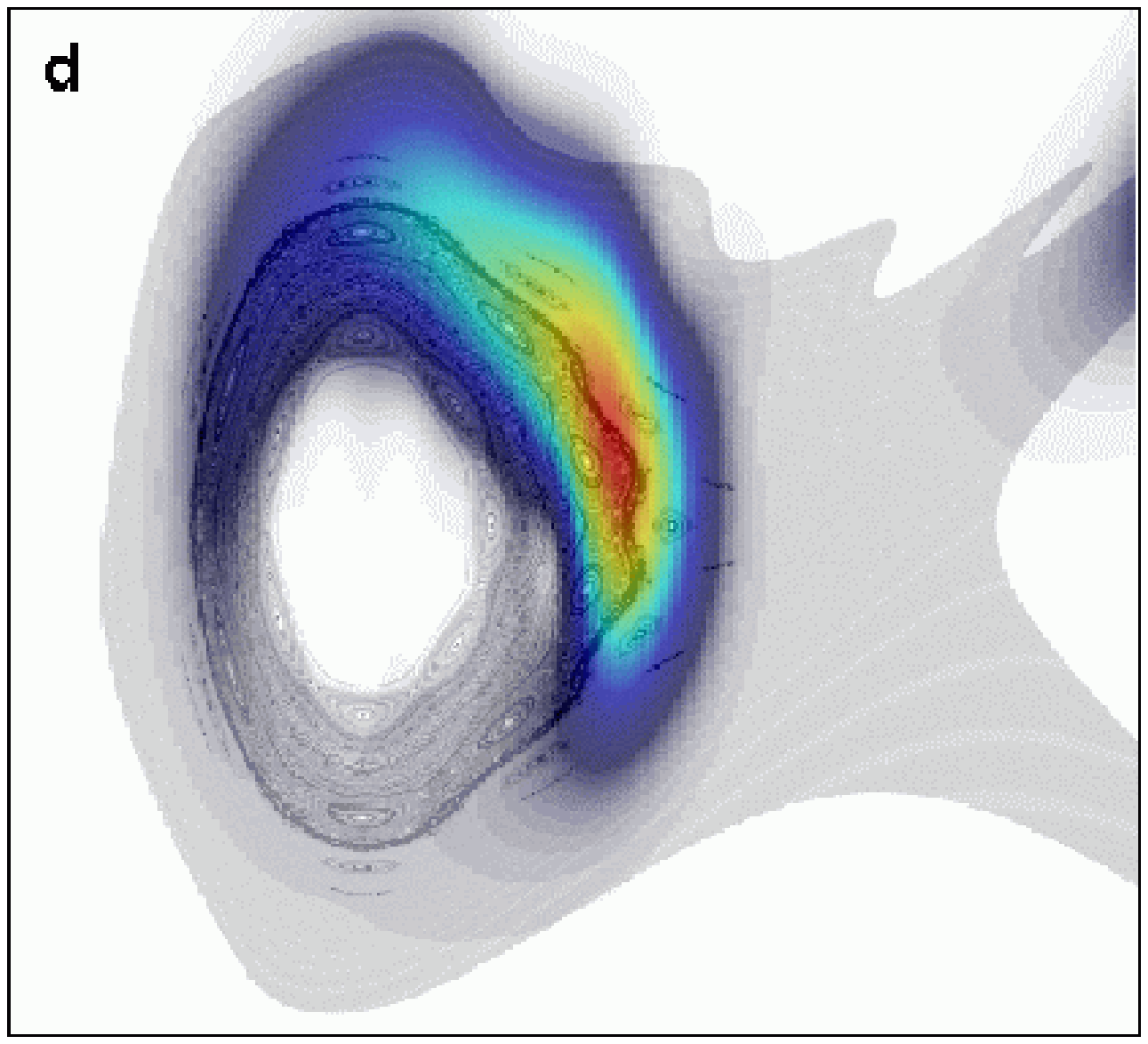}
\caption{Husimi distribution for a wave packet with $(q_{\rm
in}=100,p_{\rm in}=-1.48)$, $\sigma=2.5$, $\hbar=0.01$ and $A=0.967$
for different times. (a) $t=68$. (b) $t=73$. (c) $t=78$. (d)
$t=86$. In these plots, the colour-intensity is different; in
particular, (d) exaggerates the intensity.}
\label{fig-2}
\end{figure}

First we show in Fig.~\ref{fig-2} the Husimi distributions as a
function of time superposed to the phase portrait shown in
Fig.~\ref{fig-1}; the Husimi function is indicated by a colour-scale
code.  We choose the strength parameter $A=0.967$ and the wave packet
with $\sigma = 2.5$, $\hbar = 0.01$ and $(q_{\rm in}=100,p_{\rm
in}=-1.48)$.  At time $t=68$, the packet reaches the interaction
region, Fig.  \ref{fig-2}-a. For $t=73$, Fig. \ref{fig-2}-b, the
packet has entered the potential well near the external fixed point;
part of the packet bounces of the barrier and never enters the
well. In Fig.\ref{fig-2}-c, the probability that enters the potential
well performs its evolution along the chaotic layer of the phase
portrait at $t=78$. For $t=86$ most of the packet has left the well.
The small remaining probability gets trapped in the potential well
where it has tunneled through the surface to inner stable regions,
Fig. \ref{fig-2}-d. It is interesting to note that for $A=1$, when the
KAM-torus has became a classically penetrable cantorus the quantum
picture remains unchanged as diffusion is much slower than
tunneling. For larger values of $A$, the Husimi distribution will move
inwards to be again enclosed by the outermost KAM surface.

\begin{figure}
\onefigure[scale=0.3]{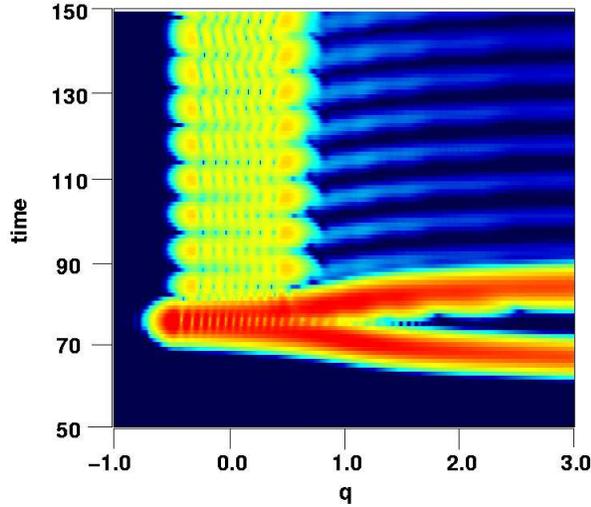}
\caption{Probability distribution in configuration space as a function
of time in a logarithmic colour-scale. After most of the probability
has been scattered, a part stays in the potential well trapped by the
stable island. This decays exponentially in echoes with a period $T
\approx 8.6$ with an estimated error of 5\%. Parameters and initial
conditions are as in Fig.  \ref{fig-2}.}
\label{fig-3}
\end{figure}

Next we show in Fig. \ref{fig-3} the probability distribution in
configuration space as a function of time in a logarithmic
colour-scale code. The parameters and initial conditions are as
above. We clearly distinguish the incoming pulse, the part directly
scattered at the barrier and another part that entered the well, but
leave directly. This direct scattering corresponds to the one shown in
Fig. \ref{fig-2}. At longer times we see an oscillating packet inside
the well with an amplitude that corresponds to the classically
forbidden region in phase space, as shown in Fig. \ref{fig-1}. Each
time the wave packet returns to the front side ({\it i.e} the largest
value of $q$) of the well it sends an ``echo'' to infinity. The time
intervals between these echoes corresponds to the oscillation times of
the packet inside the stable island which in turn corresponds to the
average classical winding time of the invariant surfaces in the region
of phase space where the packet oscillates.
Note that the winding time near the surface of the island is a generic
feature related directly to the degree of development of the horseshoe
as measured by the formal parameter $\alpha$ \cite{thesis}. The period
of rotation $T$ at the surface is given by\
\begin{equation} 
T=n+3/2\ \ \ \  {\rm if} \ \ \ \  \alpha \propto 2^{-n} \ ,
\end{equation}
with $n$ integer.  The derivation of this result will be presented
elsewhere \cite{mejia}. For $A=0.967$ we have $T \approx 10.5$ and thus
the period of rotation we see (Fig.~\ref{fig-3}), is smaller than this
estimate.  This shortening of period is not surprising and confirms
that we really see a tunneling into the stable island. As we expect,
the winding time decreases as we approach the center of the island,
(see Fig.~\ref{fig-1}).

\begin{figure}
\onefigure[scale=0.4]{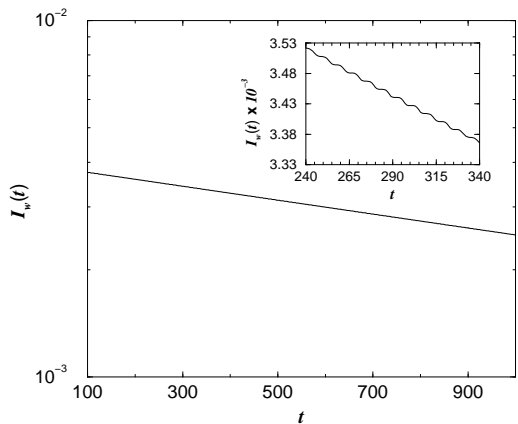}
\caption{Semi-logarithmic plot of the total intensity inside the
potential well as a function of time, $I_w(t)$. The incoming wave
packet is set as in Fig. \ref{fig-2}. In the inset, a zoom of $I_w(t)$
is shown in linear scale where the oscillations in the decay are
clearly visible.}
\label{fig-4}
\end{figure}

We evaluate the total intensity inside the potential well as a
function of time $I_w(t)$. In Fig.~\ref{fig-4} we see that the decay is
oscillatory (inset), with near constant period, but with exponentially
diminishing envelope.

\begin{figure}
\onefigure[scale=0.3]{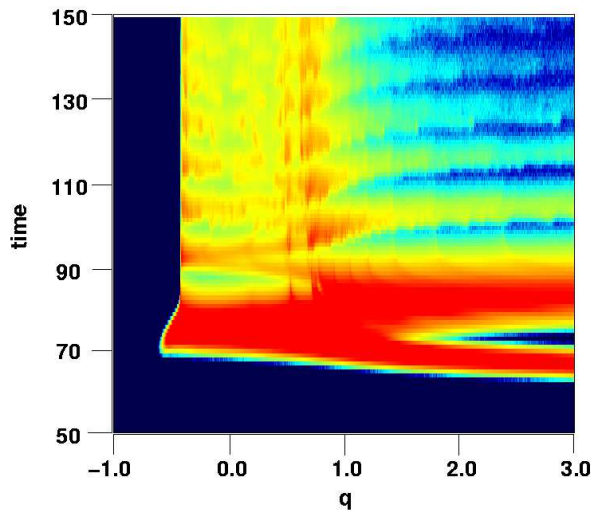}
\caption{Density distribution of classical scattered orbits as a
function of time in a logarithmic colour-scale, for $A=0.967$. The
initial conditions of the classical orbits are so, that the incoming
classical packet resembles the quantum Husimi function in phase
space. The period of the echoes is $\approx 10.9$ but the error is much
larger than in the quantum case.}
\label{fig-5}
\end{figure}

We may suspect that the echoes we see are a pure quantum phenomenon,
because of the tunneling displayed by the wave packets. Interestingly
this is not entirely true. The whole sticky fractal structure of ever
smaller islands rotates with the outer invariant surfaces of the main
island. Therefore, a small fraction of the intensity of a packet of
classically scattered particles will have a similar behaviour, though
three differences are notable: First the period is equal to the one
predicted at the surface and therefore, larger than the quantum
period, second the wave packet spreads more rapidly and third the
decay beyond the oscillations is governed by a power-law; the staying
probability decays with a power of roughly $2.554 \pm 0.005$ as is
expected for a mixed phase space \cite{karney,chirikov}.  In
Fig.~\ref{fig-5} we plot the intensity of a packet of scattered
classical particles. Note that the amplitude of the oscillation
clearly indicates that it takes place in the sticky region.

\begin{figure}
\onefigure[scale=0.4]{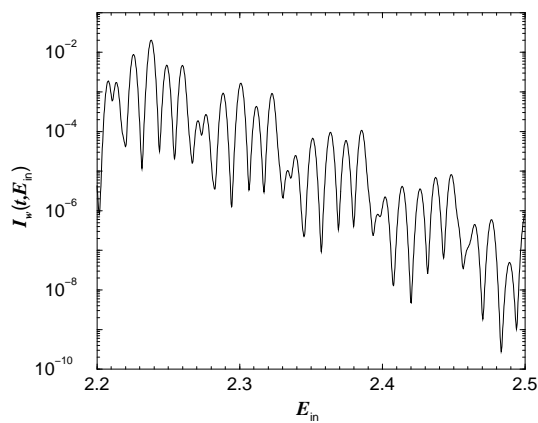}
\caption{Semi-logarithmic plot of the total intensity remaining in the
potential well at time $t=450$ as a function of the incoming
energy. This is calculated for a long pulse with $\sigma = 10$, $A=2$
and $\hbar = 0.01$.}
\label{fig-6}
\end{figure}

Returning to quantum mechanics we can also use long pulses with high
energy resolution to analyze the same phenomenon in the energy
domain. In Fig.~\ref{fig-6} we plot the total intensity remaining in
the potential well at time $t = 450$ as a function of the incoming
energy for a strength parameter $A=2$.  We clearly see the two periods
characteristic of our problem, namely the period of the pulsed system
and the one corresponding to the echoes which now is shorter as the
island is smaller.  We also calculated the $S$-matrix as a function of
the quasienergy.  It shows the same resonances we see in
Fig.~\ref{fig-6} but, a good resolution is harder to obtain.  The
presence of these sharp resonances is an additional indication that
tunneling between the regular and chaotic regions of the classical
phase portrait occurs \cite{seba}.

We have proposed the possibility to explore scattering systems
corresponding to binary and other low order horseshoes with wave and
classical scattering experiments, and find characteristic phenomena,
which we call echoes, if we use short pulses as input. From a
theoretical point of view it is interesting that experiments performed
along these lines are sensitive to the degree of development of the
corresponding horseshoe, giving us a powerful tool to explore low
developed horseshoes, where it is very hard, if not impossible, to
obtain the symbolic dynamics as pruning sets in at a very low level
\cite{rueckerl}.

From a practical point of view experiments with similar time dependent
Hamiltonians are feasible in scattering of atoms on surfaces
\cite{raizen} in the classical or almost classical domain, whereas the
corresponding experiments with electromagnetic waves seem quite
feasible with ternary horseshoes generated in appropriate cavities
\cite{richter}. Similar experiments can be performed on mesoscopic
scale.

We acknowledge useful discussions with F. Leyvraz, P. Seba and
H.-J. St\"ockmann. Financial support by DGAPA-UNAM, project IN109000
and by CONACyT, project 25-192-E is acknowledged. C.M. acknowledges a
fellowship by DGEP-UNAM.

\end{document}